\documentclass[prl,amsmath,amssymb,twocolumn,floatfix,nofootinbib]{revtex4-2}
\usepackage{color}
\usepackage[colorlinks=true,linkcolor=blue,urlcolor=blue,citecolor=blue]{hyperref}
\usepackage{graphicx}
\usepackage{times}

\usepackage{tikz}
\usetikzlibrary{arrows}

\newcommand{\Eq}[1]{Eq.~(\ref{#1})}

\newcommand{\eq}[1]{(\ref{#1})}
\newcommand{\ds}[1]{\displaystyle }

\newcommand{\bea}{\begin{eqnarray}}
\newcommand{\eea}{\end{eqnarray}}
\newcommand{\beq}{\begin{equation}}
\newcommand{\eeq}{\end{equation}}

\definecolor{mygrey}{gray}{0.5}

\definecolor{mygray}{gray}{0.4}
\newcommand{\gray}{\color{mygray}}

\newcommand{\nn}{\nonumber}
\newcommand{\ca}[1]{{\cal #1}}
\newcommand{\be}{\begin{equation}}
\newcommand{\ee}{\end{equation}}
\newcommand{\Fig}[1]{\includegraphics[width=\columnwidth]{#1}}
\newcommand{\fig}[2]{\includegraphics[width=#1\columnwidth]{#2}}

\graphicspath{{figures/},{}}

\begin{document}

\title{Lattice realization of complex CFTs: Two-dimensional Potts model with $Q>4$ states}  
  
  \author{\normalsize {\bf Jesper Lykke Jacobsen}{$^{1,2,3}$}  and {\bf Kay J\"org Wiese}{$^{1}$}}
\affiliation{{$^{1}$}CNRS-Laboratoire de Physique de l'Ecole Normale Sup\'erieure, PSL Research University, Sorbonne Universit\'e, Universit\'e Paris Cit\'e, 24 rue Lhomond, 75005 Paris, France.\\
{$^{2}$}Sorbonne Universit\'e, Ecole Normale Sup\'erieure, CNRS, Laboratoire de Physique (LPENS).\\
{$^{3}$}Institut de Physique Th\'eorique, CEA, CNRS, Universit\'e Paris-Saclay.}

\begin{abstract}
The two-dimensional $Q$-state Potts model with real couplings has a first-order transition for $Q>4$. 
We study a loop-model realization in which $Q$ is a continuous parameter. 
This model allows for the collision of a critical and a tricritical fixed point at $Q=4$, which then
emerge as complex conformally invariant theories at $Q>4$, or even complex $Q$, for suitable
complex coupling constants.
All critical exponents can be obtained as analytic continuation of known exact results for $Q \le 4$.
We verify this scenario in detail for $Q=5$ using transfer-matrix computations.
\end{abstract} 

\maketitle

The study of models in two-dimensional statistical mechanics using complex variables goes back to the   work 
of Lee and Yang \cite{LeeYang1952} on the Ising model in a complex magnetic field. Later,   complex values 
of the Ising temperature \cite{Fisher1965}, or the number of states in the Potts
model \cite{SalasSokal2001} were considered. Here  complex analysis is used to   draw conclusions about a
system for real parameters.

The $Q$-state Potts model can be reformulated as the Fortuin-Kasteleyn (FK) model \cite{FortuinKasteleyn1972}, in which each lattice bond is erased with a   probability related to the temperature. This fragments the lattice into  clusters, each with a weight $Q$
which can now  take any value. In particular, bond percolation arises for $Q \to 1$. In this formulation,
correlation functions relate to probability measures in  random geometry (clusters, hulls, backbones, {\it etc}),
which at the critical temperature form  scale- and conformally invariant fractals.

In dimension $d=2$ and   $Q$ real, criticality occurs in the range $0 \le Q \le Q_{\rm c}=4$ \cite{Baxter1973}; the phase transition turns first-order when $Q>4$, even though approximate conformal invariance remains \cite{MaHe2019}. The same happens in higher dimensions, with $Q_{\rm c}=10/3$ for $d\to 6$ \cite{WieseJacobsen2023}. 
These are examples for the annihilation of a stable (critical) fixed point (FP) with an unstable (tricritical) one upon
varying a symmetry-related continuous parameter. This phenomenon, found   long   ago \cite{CardyNauenbergScalapino1980},
 arises in  systems as different as deconfined quantum critical points \cite{WangNahumMetlitskiXuSenthil2017} and models of quantum impurity spins coupled
to a bath \cite{Nahum2022}.

It has been speculated that in the first-order regime the model may   become critical again for complex values of an unknown
set of couplings \cite{GorbenkoRychkovZan2018b,WieseJacobsen2023}. This means that the pair of FPs   does not disappear, but moves out in the complex
plane for $Q > Q_{\rm c}$. In this Letter, we show   how   this can be achieved in a specific 2d lattice Potts model.
Our model contains $Q$ as a free parameter, which can be set to any  $Q \in \mathbb C$.  
Our motivation is part of a broader scenario, with ramifications ranging from quantum field theory to probability theory,
as we now explain.

A first swathe of exact results about the 2d Potts model--and its close cousin, the ${\rm O}(n)$ model--were obtained 
in the 1980's. The lattice models were rewritten in terms of loops (the contours of the FK clusters), which in the continuum limit become level lines of a compactified bosonic
field, to which the methods of Coulomb Gas (CG) \cite{Nienhuis1984} and Conformal Field Theory (CFT)
\cite{BelavinPolyakovZamolodchikov1984} can be applied. Most of these results have since been
proved---and in some cases surpassed---by probability theorists, using methods such as Schramm-Loewner Evolution (SLE)
\cite{Cardy2005,Lawler2007} and the Conformal Loop Ensemble (CLE) \cite{Sheffield2009}.
Critical exponents (the {\it spectrum}) can be derived from field theory
\cite{DiFrancescoSaleurZuber1987}.  There are indications of a relation to Liouville CFT (LCFT) \cite{Kondev1997}, an exactly solvable interacting CFT with a continuous spectrum,
which is unitary for central charge $c \in (1,\infty)$. However, LCFT cannot   be the continuum limit of the Potts loop
model for generic $Q \in [0,4]$, since the CG for the latter is known to be non-unitary and has $c \le 1$ with a discrete spectrum \cite{DiFrancescoSaleurZuber1987}.

This situation begs the question
whether results for the loop model can be obtained from LCFT by analytic continuation through complex values of $Q$.
In LCFT the structure constant of three-point correlation functions is given by the so-called   DOZZ formula, but
analytic continuation of the latter to $c \in (-\infty,1)$ is impossible. However, a variant of LCFT, called ${\rm LCFT}_{c \le 1}$,
  exists.  
It has been established \cite{PiccoSantachiaraVitiDelfino2013,IkhlefJacobsenSaleur2016}
that the DOZZ-type formula for ${\rm LCFT}_{c \le 1}$ \cite{Zamolodchikov2005}
correctly predicts   three-point functions in the Potts loop model. In contrast, continuation of ${\rm LCFT}_{c \le 1}$
and its  
link with the Potts model for $Q > 4$ has   attracted little attention.

The authors of \cite{GorbenkoRychkovZan2018b}
proposed to analytically continue the relation between $Q$ and the CG coupling constant outside the range $Q \in [0,4]$.
They suggested that    a pair of complex CFTs describe the $Q = 4 + \epsilon$ Potts
model to first order in $\epsilon > 0$.
Our   numerical study of the lattice model makes this link precise
and reveals that the correspondence  extends not only to $Q>4$, but   to a large
portion of the complex $Q$-plane. In particular, we provide   evidence that the spectra of the
two complex CFTs are the appropriate analytic continuation of those   \cite{DiFrancescoSaleurZuber1987}
for the ${\rm LCFT}_{c \le 1}$ loop models.
We expect recent results on three- and four-point correlation functions \cite{HeJacobsenSaleur2020,GransSamuelssonJacobsenNivesvivatRibaultSaleur2023,NivesvivatRibaultJacobsen2023}
and the symmetries of the space of states \cite{JacobsenRibaultSaleur2023} of the loop models
to carry over to the complex CFT as well.

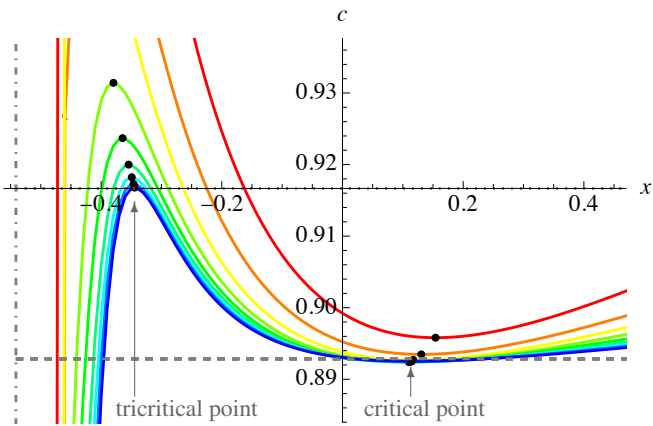
\begin{figure}[b]
{\setlength{\unitlength}{1mm}\begin{picture}(86,56)
\put(0,0){\Fig{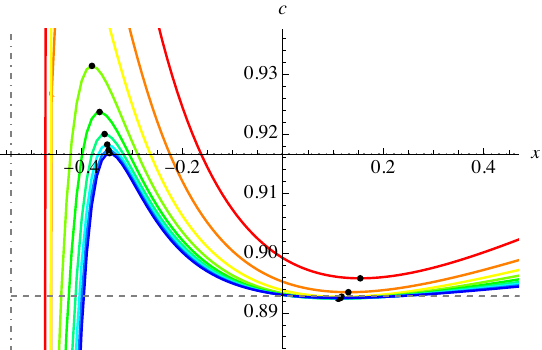}}
\put(48,1){\gray critical point}
\put(54.2,3.6){\gray\vector(0,1){4}} 
\put(15,1){\gray tricritical point}
\put(17.5,3.6){\gray\vector(0,1){26}} 
\end{picture}}
\caption{$c_L(x)$ for $b=8$, $Q \simeq 3.41421$, using sizes $L=4$ (red) to $12$ (blue). The discontinuity near $x=-1/\sqrt{Q}$
(vertical dot-dashed line) is related to level crossings for $K_2 \to -\infty$.
These prevent us from using this approach for smaller $Q$.
CFT predictions are shown for $B_8$ (dashed line) and $B_{-8}\equiv B_9$ (axis, and gray dots). 
}
\label{f:B8}
\end{figure}
\begin{figure}[t]
\centerline{\fig{0.5}{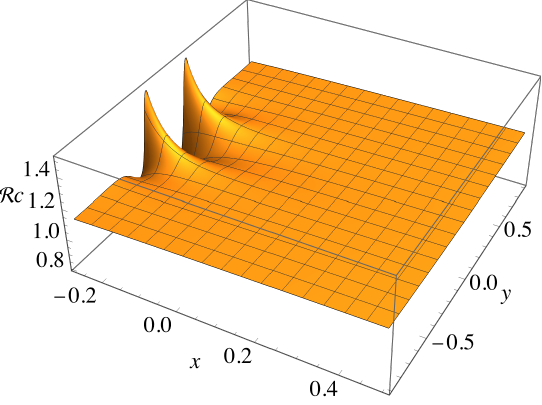}\fig{0.5}{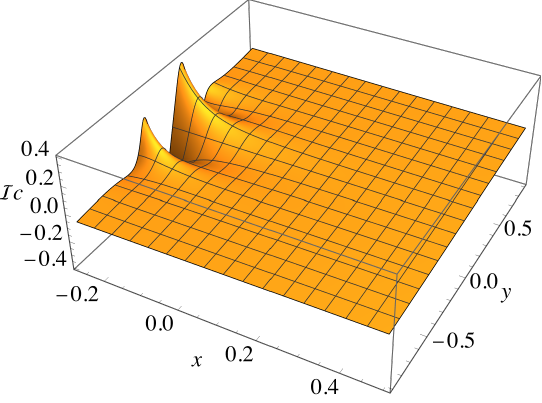}}
\caption{$Q=5$, $L=9$, selected  data for $\Re c$ (left) and $\Im c$ right, as a function of $z=x+iy$.}
\label{f:Q=5}\centerline{\fig{0.5}{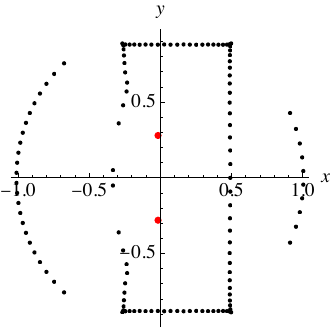}\fig{0.5}{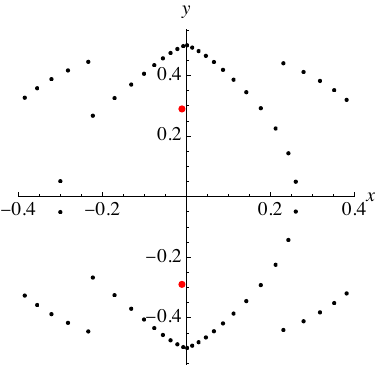}}
\caption{Zeros of $c'(z)$. Left: $L=10$ using measures on a rectangular grid of points. Right: $L=13$ using measures only   along the imaginary axis.}
\label{f:zerosQ=5}
\end{figure}
Our starting point is a $Q$-state Potts model on a triangular lattice with nearest-neighbor interactions $K_2$, and a three-spin
interaction $K_3$ on each up-pointing triangle. Making an FK expansion  
\cite{WuLin1980}, one obtains a loop model
on a shifted triangular lattice, with five possible diagrams at each vertex. These are, with 
circles denoting the loci of the   Potts spins, 
$$
\begin{tikzpicture}
 \draw (0.0,0.0) -- (1.0,0.0) -- (0.5,0.866) -- cycle;
 \draw [fill] (0,0) circle (.5ex);
 \draw [fill] (1.0,0.0) circle (.5ex);
 \draw [fill] (0.5,0.866) circle (.5ex);
 \draw [thick,red,rounded corners=2mm] (-0.166,0.289) -- (0.5,0.289) -- (0.167,-0.289);
 \draw [thick,red,rounded corners=2mm] (0.167,0.866) -- (0.5,0.289) -- (0.833,0.866);
 \draw [thick,red,rounded corners=2mm] (1.166,0.289) -- (0.5,0.289) -- (0.833,-0.289);
 \draw (0.5,-0.3) node {$1$};
\end{tikzpicture} \quad
\begin{tikzpicture}
 \draw (0.0,0.0) -- (1.0,0.0) -- (0.5,0.866) -- cycle;
 \draw [fill] (0,0) circle (.5ex);
 \draw [fill] (1.0,0.0) circle (.5ex);
 \draw [fill] (0.5,0.866) circle (.5ex);
 \draw [thick,red,rounded corners=2mm] (1.166,0.289) -- (0.5,0.289) -- (0.833,0.866);
 \draw [thick,red,rounded corners=2mm] (0.167,0.866) -- (0.833,-0.289);
 \draw [thick,red,rounded corners=2mm] (0.167,-0.289) -- (0.5,0.289) -- (-0.166,0.289);
 \draw (0.5,-0.3) node {$z$};
\end{tikzpicture} \quad
\begin{tikzpicture}
 \draw (0.0,0.0) -- (1.0,0.0) -- (0.5,0.866) -- cycle;
 \draw [fill] (0,0) circle (.5ex);
 \draw [fill] (1.0,0.0) circle (.5ex);
 \draw [fill] (0.5,0.866) circle (.5ex);
 \draw [thick,red,rounded corners=2mm] (-0.166,0.289) -- (0.5,0.289) -- (0.167,0.866);
 \draw [thick,red,rounded corners=2mm] (0.167,-0.289) -- (0.833,0.866);
 \draw [thick,red,rounded corners=2mm] (0.833,-0.289) -- (0.5,0.289) -- (1.166,0.289);
 \draw (0.5,-0.3) node {$z$};
\end{tikzpicture} \quad
\begin{tikzpicture}
 \draw (0.0,0.0) -- (1.0,0.0) -- (0.5,0.866) -- cycle;
 \draw [fill] (0,0) circle (.5ex);
 \draw [fill] (1.0,0.0) circle (.5ex);
 \draw [fill] (0.5,0.866) circle (.5ex);
 \draw [thick,red,rounded corners=2mm] (0.167,0.866) -- (0.5,0.289) -- (0.833,0.866);
 \draw [thick,red,rounded corners=2mm] (-0.166,0.289) -- (1.166,0.289);
 \draw [thick,red,rounded corners=2mm] (0.167,-0.289) -- (0.5,0.289) -- (0.833,-0.289);
 \draw (0.5,-0.3) node {$z$};
\end{tikzpicture} \quad
\begin{tikzpicture}
 \draw (0.0,0.0) -- (1.0,0.0) -- (0.5,0.866) -- cycle;
 \draw [fill] (0,0) circle (.5ex);
 \draw [fill] (1.0,0.0) circle (.5ex);
 \draw [fill] (0.5,0.866) circle (.5ex);
 \draw [thick,red,rounded corners=2mm] (0.167,-0.289) -- (0.5,0.289) -- (0.833,-0.289);
 \draw [thick,red,rounded corners=2mm] (0.833,0.866) -- (0.5,0.289) -- (1.166,0.289);
 \draw [thick,red,rounded corners=2mm] (-0.166,0.289) -- (0.5,0.289) -- (0.167,0.866);
 \draw (0.5,-0.3) node {$1$};
\end{tikzpicture}
$$
Taking equal weights of the first and last diagram imposes a relation between $K_3$ and $K_2$ which ensures self-duality
\cite{WuLin1980}. The three middle diagrams have weight $z = ({\rm e}^{K_2} - 1) / \sqrt{Q}$, and we   write this
as $z = x + i y$ with $x, y \in \mathbb{R}$. In addition to these local weights, there is a non-local
factor $n = \sqrt{Q}$ for each red loop.

We study this model via the transfer matrix $T_L$ which builds a row of $L$ triangles, with
periodic boundary conditions. Thus $T_L$ propagates $\sqrt{3}/2$ lattice spacings upwards and $1/2$ to  the right.
The operator $\check{R}_k$ that propagates through triangle number $k$ is  
\bea
 \check{R}_k = e_{2k-1} + z \big( e_{2k} e_{2k-1} + 1 + e_{2k-1} e_{2k} \big) + e_{2k} ,
\eea
where $e_i$ are generators of the periodic Temperley-Lieb algebra \cite{JacobsenRibaultSaleur2023} on $2L$ sites.
Each of the five terms corresponds to one of the above diagrams.
We have $T_L = u^{-1} \text{qTr} \check{R}_L \cdots \check{R}_2 \check{R}_1$, where $\text{qTr}$ denotes the quantum trace over the
horizontal space, and $u$ shifts the sites cyclically towards the right. We  diagonalize $T_L$
in the space of link patterns, sometimes called standard modules ${\cal W}_{j,z^2}$, with $2j$ defect lines (FK cluster boundaries)
propagating from bottom to top, and (pseudo) momentum $z$ describing the winding of lines with respect to the periodic boundary
condition; see \cite{JacobsenRibaultSaleur2023} for details.

The effective central charge $c_L$ and critical exponents $\Delta_L$ are obtained   \cite{BloteCardyNightingale1986} from the finite-$L$ corrections
of the leading eigenvalues of $T_L$. For $c_L$ we need two consecutive sizes, $L$ and $L+1$.
We use the ground-state sector, in which $T_L$ acts on defect-free link patterns $(j=0)$.

For $Q\le 4$, the model contains a critical and a tricritical point, obtained for fine-tuned values of $x$. 
Fig.~\ref{f:B8} shows $c_L$ as a function of $x$: 
  $c_L$ is minimal at the critical point, and maximal at the tricritical point.
The values at these points agree   with the predictions of CFT,
\bea
\label{Q-and-c-of-b}
Q &=& 4 \cos\!\Big(\frac{\pi}b\Big)^2, \quad
c(t) = 1-\frac{6}{t(t-1)}.
\eea
Here $Q$ defines the parameter $b$, and we take $t = b>0$ for the critical point and $t = -b$ for the tricritical one.
The corresponding CFTs are denoted $B_b$. They are minimal models \cite{BelavinPolyakovZamolodchikov1984} for $b$ rational,
but the loop model describes generic values of $b$ (which are real for $Q \le 4$).
We   identify $B_{-b}$ with $B_{b+1}$. Conformal weights are parametrized by the Kac formula
\bea
h_{r,s}^t&=& \frac{[t (r-s)+s]^2-1}{4 (t-1) t},
\eea
and below we   write $\Delta_{r,s} = 2 h_{r,s}^t$ for the corresponding critical exponents.

{\begin{figure*}[t]\newcommand{\here}{0.79}
\leftline{{\parbox{\here\columnwidth}{\fig{\here}{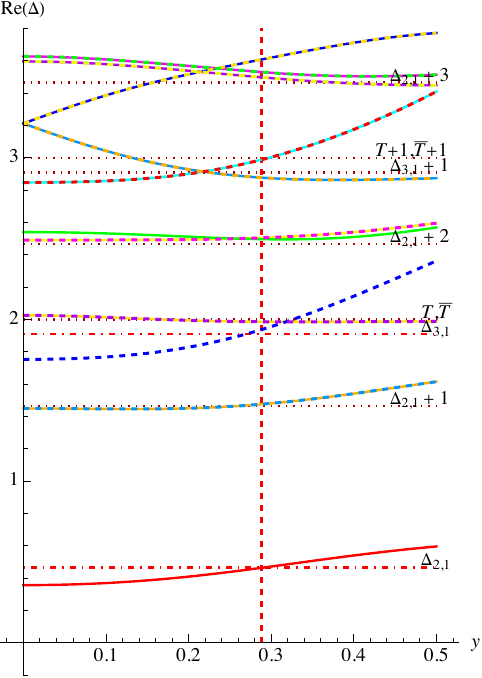}}}~~~{\parbox{\here\columnwidth}{\fig{\here}{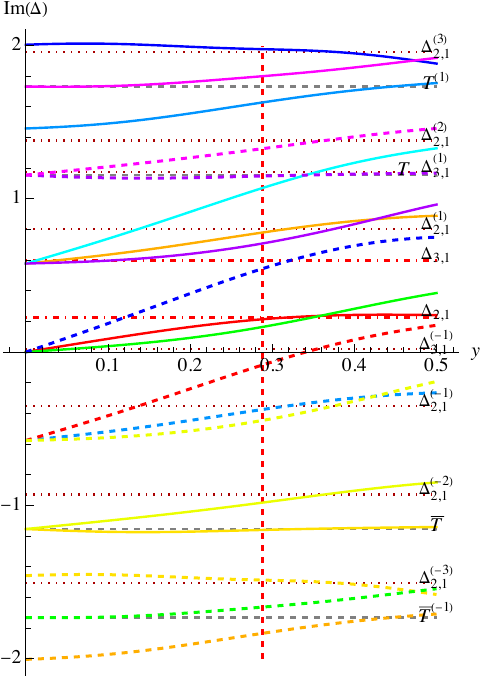}}}
~~~{\parbox{3.8cm}{\begin{tabular}{|c|c|r|}
\hline
state & operator & spin  \\
\hline
1 & $\mbox{GS}=\Phi_{1,1}$ & 0  \\
2 & $\epsilon= \Phi_{2,1}$ & 0  \\
3 & $L_{0,-1} \Phi_{2,1}$ & $-1$  \\
4 & $ L_{-1,0} \Phi_{2,1}$ & $1$   \\
5 & $\epsilon'=\Phi_{3,1}$ & 0  \\
6 & $\overline T \equiv L_{0,-2} \Phi_{1,1}$ & $-2$  \\
7 & $ T \equiv L_{-2,0} \Phi_{1,1}$ & $2$  \\
8 & $L_{0,-2} \Phi_{2,1}$ & $-2$  \\
9 & $L_{-2,0} \Phi_{2,1}$ & $2$  \\
10 & $L_{-1,-1} \Phi_{2,1}$ & $0$   \\
11 & $L_{0,-1} \Phi_{3,1}$ & $-1$   \\
12 & $L_{-1,0} \Phi_{3,1}$ & $1$ \\
13 & $L_{0,-1} \overline T$ & $-3$ \\
14 & $L_{-1} T$ & $3$  \\
15 & $L_{0,-3} \Phi_{2,1}$ & $-3$  \\
16 & $L_{-3,0} \Phi_{2,1}$ & $3$  \\
17 & $L_{-1,-2} \Phi_{2,1}$ & $-1$ \\
18 & $L_{-2,-1} \Phi_{2,1}$ & $1$  \\
19 & $L_{0,-3} \Phi_{2,1}$ & $-3$  \\
20 & $L_{-3,0} \Phi_{2,1}$ & $3$   \\
\hline
\end{tabular}}}}
\caption{Real and imaginary part $\Delta_{r,s}^{(S)}$ of spectrum in the ground-state sector for $Q=5$, $L=14$, taking only the first 20 EVs. 
Primaries in red, dashed, descendents in darker red, dotted. The red dashed vertical lines denote the position of the critical point.
For details, see \cite{JacobsenWieseInPreparation2024}.}
\label{f:SpectrumQ=-5}
\end{figure*}}
Fig.~\ref{f:B8}   shows that upon increasing $L$, $c_L(z)$  around the critical point becomes shallower, while around the 
tricritical point it becomes more pointed. A finite-size analysis of the curvature $c''(z)$ allows us to extract the correction-to-scaling exponent $\omega$ via 
\be
c_L''(z_c) \sim L^{-2 \omega}. 
\ee%
It is related to the dimension $\Delta$ of the perturbing operators as
\bea\label4
\!\!\!\Delta= 2+\omega^{\rm crit} \approx 2.57  , &\quad& \Delta= 2+\omega^{\rm tricrit} \approx  1.55~~ \\
\Delta_{3,1}^{\rm crit} =18/7= 2.5714, &\quad & \Delta^{\rm tricrit}_{3,1}=14/9=1.5556. ~~~
\label5
\eea
We find good agreement of measurements \eq{4} with the predictions of CFT \eq5.
To summarize: we are   able to identify a critical or tricritical point by   $c'(z)=0$, and   use a finite-size analysis on $c_L''(z_c)$ to decide whether it is attractive or repulsive. Such a prescription is   missing in related work on the ${\rm O}(n)$ model, where the location of the critical point is known analytically \cite{HaldarTavakolMaScaffidi2023}.

To proceed, we enlarge our transfer-matrix treatment to complex values of $z$. The crucial observation is that for all $L$ and $Q$,  $c_L(z)$ is an {\em analytic function} of $z$. It   is  well represented by a   
polynomial in $z$ of moderate order (30-100), 
with a negligible error ($\ll 10^{-4}$). This is valid for $Q\le 4$ and $Q>4$:
a plot for $Q=5$ is shown in Fig.~\ref{f:Q=5}. 
The critical points are obtained by solving the polynomial  $c_L'(z)=0$, which can be done with high precision. The result is shown on the left panel of Fig.~\ref{f:zerosQ=5}. 
There is a pair of critical points next to the imaginary line (in red), and spurious zeros on the boundary of the domain of convergence (in black). The latter roughly indicate the size of the   grid of measurement points in Fig.~\ref{f:Q=5}.
Analytic continuation is so powerful that one can   replace the grid by points  on the imaginary axis only,
see the right panel of Fig.~\ref{f:zerosQ=5}. This drastically reduces the work to be done. 
 Extrapolating to $L=\infty$, we find 
\be\label{c-Q=5-measured}
c^{\rm num}_{Q=5} \approx 1.1377 + 0.0221 i ,
\ee
in remarkable agreement with the CFT prediction, \Eq{Q-and-c-of-b}:
\be\label{c-Q=5-theory}
c^{\rm CFT}_{Q=5} = 1.13755 + 0.0210687 i .
\ee
This firmly establishes that there is a {\em complex CFT}, and that it can be identified from transfer-matrix calculations. 

We can go further and obtain the full spectrum of this complex CFT.
We study first the spectrum in the ground-state sector,
using the quotient representation $\overline{\mathcal W}_{0,\mathfrak{q}^2}$ of defect-free link patterns of dimension
$\tfrac{1}{L+1} {2L \choose L}$, where $n = \mathfrak{q} + \mathfrak{q}^{-1}$; see \cite{JacobsenRibaultSaleur2023} for details.
Fig.~\ref{f:SpectrumQ=-5} shows the exponents corresponding to the first 20 eigenvalues (EV) in this sector.
As the critical points have $x \approx 0$ (see Fig.~\ref{f:zerosQ=5}), we take $x=0$ and plot as a function of $y \ge 0$.
The fixed point is indicated by the vertical red dashed line. 

This spectrum contains the spinless primary operators $\Phi_{r,1}$ with eigenvalues $\Delta_{r,1}= 2 h_{r,1}^{-b}$, 
starting with $r=1$ for the vacuum (baseline). We verify the appearance of $\Phi_{2,1}$ and $\Phi_{3,1}$,
with the proper dimensions both for the real and imaginary parts.  
Descendents have a non-vanishing spin $S$. The dimension of the  primary increases by $1$ for each level, and $S/\sqrt3$
for the imaginary part. The latter arises since $T_L$ propagates by $\sqrt{3/4}$ upwards, and $1/2$   rightwards,
with ratio $\sqrt 3$. Using $L_{-N,-\bar{N}} \Phi_{r,1}$ as a shorthand for any descendent on chiral level $N$ and antichiral
level $\bar{N}$, we get
\be
\Delta (L_{-N,-\bar{N}} \Phi_{r,1} )  = \Delta_{r,1} + (N+\bar{N}) + \tfrac{1}{\sqrt{3}} (N-\bar{N})  i .
\ee
The $\Phi_{r,1}$ form Kac modules with one  null state  on level $r$. The content of primaries and the structure of the modules
were predicted for $Q < 4$   \cite{DiFrancescoSaleurZuber1987} and numerically verified for a corresponding loop model
\cite{JacobsenSaleur2019}. We   see in Fig.~\ref{f:SpectrumQ=-5} that 
all of this is   perfectly respected  at $Q=5$ for the 20 lowest-lying states: 
 we conjecture that the results for the spectrum carry over to the complex CFT by analytic continuation.
Generalizations to sectors with 
defect lines will be discussed elsewhere \cite{JacobsenWieseInPreparation2024}. 

From finite-size scaling we find $\omega$, which satisfies $2+\omega \approx\Delta_{3,1}= 2 h_{3,1}^{-b}$, 
\be\label{omega-Q=5}
\omega = -0.12(2) + 0.62(2) i, \quad \Delta_{3,1}=1.9083 + 0.598652 i. \nonumber
\ee

\begin{figure}
{\fig{0.5}{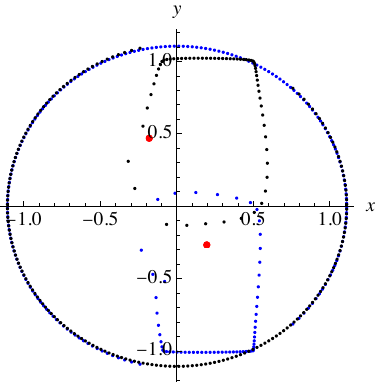}}{\fig{0.5}{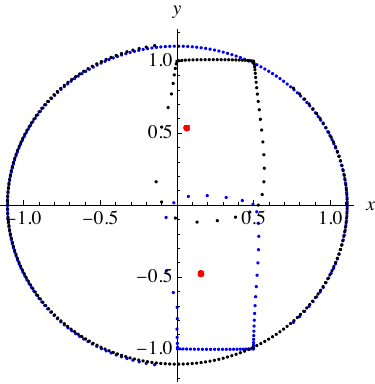}}
\caption{Zeros of $c'(z)$ at $L=9$. Left: $Q=5+2i$. Right: $Q=8+i$. We have used separate fits for the upper (black) and lower (blue) half planes.}
\label{f:zeros-complex-Q}
\end{figure}

We now proceed to complex values of $Q$. In Fig.~\ref{f:zeros-complex-Q} we show the location of the zeros of $c_L'(z)$ evaluated in the plane for $Q=5+2i$ and $Q=8+i$. 
Since $Q^*\neq Q$ they are no longer complex conjugate of each other, and   the two fixed points are distinct.  
We can again measure the central charge and compare to the theoretical prediction, first for $Q=5+2i$,  
\be
\begin{array}{rclcrcl}
c_{\rm upper}^{\rm num} &\approx&  1.108 +0.252 i, &\quad& c_{\rm lower}^{\rm num} &\approx& 1.247 +0.228 i,
\\
c(-b_Q) &=& 1.1122 + 0.2505 i,&\quad &
c(b_Q) &=& 1.2464 + 0.2293 i . \nn
\end{array}
\ee
For $Q=8+i$ we find
\be
\begin{array}{rclcrcl}
c_{\rm upper}^{\rm num} &\approx& 1.407 +0.207 i  , &\quad& c_{\rm lower}^{\rm num} &\approx&  1.479 -0.047 i,
\\
c(-b_Q) &=&  1.4073 +0.2041  i,&\quad &
c(b_Q) &=& 1.4778 -0.0420 i.  \nn
\end{array}
\ee%
Both cases are in good agreement. 
Our procedure continues to work for $Q=10$, $Q=20$ and $Q=40$. 

\begin{figure}[!t]
\centerline{{\fig{0.93}{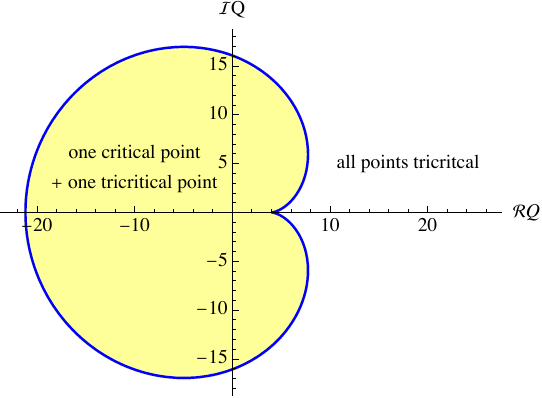}}} 
\caption{In yellow, the region where $\Re \Delta_{3,1}(b)>2$. In blue, the parametric curve \eq{parametric-curve-PRL}.}
\label{f:transition-critical-tricritical}
\end{figure}

The reader may wonder when the complex fixed points found here are critical, or tricritical. We have shown that for $Q<4$ and $Q=5+2i$   both a critical and a tricritical point exist, while for $Q=5$ and $Q=8+i$ both points are tricritical. 
In general, this is determined by the values of $\Delta_{3,1}$.   
Inside the    yellow-shaded region on  Fig.~\ref{f:transition-critical-tricritical}, one of them is irrelevant  and one relevant, whereas outside both are relevant. 
The boundary is given by   
\be\label{parametric-curve-PRL}
Q(\phi)=2+2 \cos \Big(\frac{2\pi }{1+i \phi}\Big),\quad \phi \in \mathbb R.
\ee
Let us finally draw the two critical points for complex $Q$, with $\Re Q>4$, as well as their two complex conjugates, which are critical points for $Q^*$. Moving $Q$ to the real axis, the four   points   merge in pairs, and only two fixed points remain.

Our model can be reformulated in terms of non-Hermitian quantum mechanics.
To this end, we take an anisotropic limit of the transfer matrix, by stretching infinitely along the direction of propagation (with a
shift towards the right). Up to a constant and rescaling $z$ one finds the
Hamiltonian \cite{IkhlefJacobsenSaleur2009}
\be 
\ca H = \sum_{i=1}^L    e_i   + \tilde z  \left( e_i e_{i+1} + e_{i+1} e_i\right).
\ee
A previous study of this quantum spin chain   found, for real loop weight $\sqrt Q = n \in [n^*,2)$ with $n^* \simeq 1.5$, a critical phase terminated
by a tricritical end point. 
By universality, we posit that the spin chain     exhibits complex critical points for $n > 2$, and more
generally $n \in \mathbb{C}$. This can be studied by using the XXZ representation of the $e_i$ generators
\cite{JacobsenWieseInPreparation2024}.
For integer $Q$, the results should carry over   to the quantum chain formulated in terms of $Q$-component Potts spins,
but   finer details (such as multiplicities and operator product expansions) may differ due to the change of
representation.

\acknowledgements
We thank Y.-C.~He, A.~Ludwig, A.~Nahum, S.~Ribault, S.~Rychkov, M.~Salmhofer and H.~Saleur for stimulating discussions. 
This work was supported by the French Agence Nationale de la Recherche (ANR) under grant ANR-21-CE40-0003 (project CONFICA), 
and in part by grant NSF PHY-2309135 to the KITP.

\ifx\doi\undefined
\providecommand{\doi}[2]{\href{http://dx.doi.org/#1}{#2}}
\else
\renewcommand{\doi}[2]{\href{http://dx.doi.org/#1}{#2}}
\fi
\providecommand{\link}[2]{\href{#1}{#2}}
\providecommand{\arxiv}[1]{\href{http://arxiv.org/abs/#1}{#1}}
\providecommand{\hal}[1]{\href{https://hal.archives-ouvertes.fr/hal-#1}{hal-#1}}
\providecommand{\mrnumber}[1]{\href{https://mathscinet.ams.org/mathscinet/search/publdoc.html?pg1=MR&s1=#1&loc=fromreflist}{MR#1}}



\end{document}